%% file: main.tex
\documentclass[conference]{IEEEtran}

\usepackage[numbers,sort&compress]{natbib}
\usepackage{amsmath,amssymb,amsfonts}
\usepackage{booktabs}
\usepackage{siunitx}
\usepackage{pifont}
\usepackage{algorithmic}
\usepackage{graphicx}
\usepackage{textcomp}
\usepackage{xcolor}
\usepackage{url}
\def\BibTeX{{\rm B\kern-.05em{\sc i\kern-.025em b}\kern-.08em
    T\kern-.1667em\lower.7ex\hbox{E}\kern-.125emX}}
\usepackage{pifont}
\newcommand{\cmark}{\ding{51}}%
\newcommand{\xmark}{\ding{55}}%

\newcommand{\fix}[1]{\textcolor{black}{#1}}

\newcommand{\camred}[1]{\textcolor{black}{#1}}

\begin{document}

\title{Efficient and Fast Generative-Based Singing Voice Separation using a Latent Diffusion Model}

\author{\IEEEauthorblockN{Genís Plaja-Roglans\IEEEauthorrefmark{1}\IEEEauthorrefmark{2}, Yun-Ning Hung\IEEEauthorrefmark{1}, Xavier Serra\IEEEauthorrefmark{2}, and Igor Pereira\IEEEauthorrefmark{1}}
\IEEEauthorblockA{\IEEEauthorrefmark{1}Music.AI, Salt Lake City, Utah, United States}
\IEEEauthorblockA{\IEEEauthorrefmark{2}Music Technology Group, Universitat Pompeu Fabra. Barcelona, Spain}
}

\maketitle

\begin{abstract}
Extracting individual elements from music mixtures is a valuable tool for music production and practice. 
%
While neural networks optimized to mask or transform mixture spectrograms into the individual source(s) have been the leading approach, the source overlap and correlation in music signals poses an inherent challenge. Also, accessing all sources in the mixture is crucial to train these systems, while complicated. 
%
%
%
%
%
Attempts to address these challenges in a generative fashion exist, however, the separation performance and inference efficiency remain limited. In this work, we study the potential of diffusion models to advance toward bridging this gap, focusing on generative singing voice separation relying only on corresponding pairs of isolated vocals and mixtures for training.
To align with creative workflows, we leverage latent diffusion: the system generates samples encoded in a compact latent space, and subsequently decodes these into audio. This enables efficient optimization and faster inference.
%
Our system is trained using only open data.
%
\fix{We outperform existing generative separation systems, and level the compared non-generative systems on a list of signal quality measures and on interference removal. We provide a noise robustness study on the latent encoder, providing insights on its potential for the task.} \camred{We release a modular toolkit for further research on the topic.}\footnote{Please see \url{https://github.com/WeAreMusicAI/dmx-diffusion}}
\end{abstract}

\section{Introduction}\label{sec:intro}
Deep generative audio modeling has emerged as a widely explored topic, with important advances specially attributed to denoising diffusion probabilistic models (DDPM)~\cite{diffusion-first}.
These generative systems have demonstrated impressive performance for creative purposes~\cite{schneider2023mousai,evans2024fast}.
By introducing stringent conditioning techniques~\cite{universe_serra, controlnet}, the generative potential of DDPM may be used to address audio inverse problems~\cite{universe_serra, moliner_inverse, mariani2023multi}, showing promise for multiple applications crucial to music creation. Few examples are audio or speech enhancement~\cite{universe_serra}, upsampling~\cite{diffusion-upsampling}, and more recently source separation~\cite{mariani2023multi, msdm_ldm}.

Music source separation (MSS) involves isolating individual elements from a musical mixture~\cite{unet}. It plays an important role in music creation, practice, and analysis~\cite{plaja2023carnatic}. This task is generally addressed via neural networks that mask or transform the spectrogram of a mixture to extract the individual sources~\cite{sdx_2023}. However, these face an inherent challenge due to the significant overlap
between musical sources, which may limit performance.
Additionally, synthesizing the estimated spectrograms into the time domain introduces further complexity~\cite{unet}, and predicting the phase of complex spectrograms is a studied but challenging task~\cite{choi2020cac}. Moreover, having access to all sources that linearly sum up to the mixture is crucial to train these systems~\cite{kim2021kuielab}, but acquiring such data is costly~\cite{instglow}.

These challenges are crucial as music practitioners value high-quality, clean separations. While recent deterministic models have achieved impressive performance on objective separation metrics~\cite{ht_demucs, lu2024music}, it remains unclear whether these metrics fully-capture perceptual quality~\cite{cano2016evaluation, guso2022loss}. 
%
This is more pronounced for generative models, which inherently sample from a modeled data distribution. This often results in outputs with minor, potentially imperceptible deviations from the target when addressing inverse problems.
These subtle differences are disproportionately penalized by the separation metrics~\cite{gan_prior, pritish_sep}. However, users may prioritize perceptual quality and cleanliness over an exact copy of the target signal.

Despite their generation potential, the computational cost of training DDPM remains a challenge, while large datasets are normally required~\cite{diffusion-first, universe_serra}.
To alleviate this, latent diffusion models (LDM) were proposed~\cite{rombach2021highresolution}.
%
These systems are trained to generate samples encoded in a learned and compact latent representation which is leveraged from an autoencoder optimized for the target data. 
Thereby, latent diffusion enables faster and more efficient optimization. More importantly, inference can be run effectively with less computing resources, which is crucial to bring these tools to music practitioners.

%
We explore the potential of latent diffusion to separate the singing voice, a crucial but complex source, having solely access to solo vocal tracks and the corresponding mixtures for training.
Recently, DDPM have been employed to separate musical sources both in the time~\cite{mariani2023multi} and latent domains~\cite{msdm_ldm}. However, these studies primarily focus on synthetic instrumental mixtures and often exclude vocals, limiting their applicability to real-world recordings.
To address this, we leverage the latent space of EnCodec, a high-fidelity neural audio codec~\cite{defossez2022highfi}, to train a latent diffusion model that generates encoded vocal representations conditioned on music mixtures.

The objective and perceptual experiments indicate that the proposed system outperforms the generative baselines, and levels the compared non-generative systems on a set of signal quality metrics and interference removal, albeit the potential presence of high-frequency generation and decoding artifacts leave room for further work.
The latent diffusion framework enhances efficiency and facilitates bringing generative tools to music creators.
%
%
We perform a noise robustness experiment on EnCodec,
%
%
and publish a modular Python toolkit to reproduce our work and perform further research on the topic.

\section{Related work}\label{sec:related_work}
%

\textbf{Generative-based MSS.}
%
\fix{Attempts to address MSS using generative modeling have been done, using flow-based generators~\cite{instglow}, generative adversarial networks (GAN)~\cite{gan_prior}, or variational autoencoders (VAE)~\cite{mancusi2021unsupervised}. 
Previous to these, in~\cite{pritish_sep} the authors estimated vocoder features from mixture signals to synthesize the corresponding vocals.
Yet, these systems have not shown competitive performance against non-generative separation systems, and are usually limited by the computational cost.
More recently, DDPM have demonstrated potential for inverse audio problems and have been applied to MSS, marking an improvement over previous generative separation approaches~\cite{mariani2023multi}.} Currently, diffusion models are regarded as the most promising generative technology for audio source separation given its generalization capability~\cite{araki20253030years}.

\textbf{DDPM for separating musical sources.} While few separation attempts using diffusion have been done~\cite{diff-source-sep-only,scheibler2022diffusionbased,yu2023zeroshot,mariani2023multi,msdm_ldm}, none of these address singing voice separation from music accompaniments.
The multi-source diffusion model (MSDM)~\cite{mariani2023multi} is trained to model the joint probability of musical sources to perform generation, impaiting, and separation simultaneously. More recently, an implementation of this model that operates on the latent space of a pre-trained variational autoencoder has been proposed~\cite{msdm_ldm} under the name of MSG-LD. However, the vocal source is not included in the main study of these works, which are trained on synthetic instrumental data~\cite{slakh}.
In the work we focus on source-only vocal separation, addressing a different aspect of music production, and providing users with tools for vocal-specific analysis. Conveniently, DDPM have demonstrated capability to generate singing vocals in the latent space~\cite{hwang2023hiddensinger}, but not yet to disentangle this important source from correlated musical instruments. 


\textbf{Neural audio codecs.} Diverse autoencoder classes have been used to implement latent diffusion~\cite{rombach2021highresolution, hwang2023hiddensinger, demerle2024latent}. Recently, neural audio coding architectures have also been demonstrated useful~\cite{evans2024fast}.
These systems provide enormous audio signal compression and perceptually-optimized latent spaces~\cite{defossez2022highfi, Kumar2023HighFidelityAC}.
Interestingly, the latent space of a neural codec has been disentangled to perform audio source separation~\cite{bie2025sdcodec}, albeit not in separating singing vocals from the corresponding accompaniment.
While diffusion models have been used to directly decode EnCodec discrete tokens to audio signals~\cite{san2023discrete}, in this work we focus on creating these latent vectors for unseen cases. Moreover, cascading diffusion processes may slow down the entire process and increase the computational expense, making the system less suitable for music creators.

While the autoencoder may be pre-trained on the same data used to train the diffusion model~\cite{hwang2023hiddensinger,demerle2024latent}, here we study the feasibility of bypassing this step. Thus, we aim to reduce the dependency of the system on the training data and streamline the development process boosting its efficiency. 

\section{Method}\label{sec:method}
\subsection{Diffusion forward process}\label{sec:diffusion}
A diffusion process is defined by a Markov chain of $T$ steps that converts given a sample 
$y_{\sigma_{0}} \in \mathbb{R}$ pertaining to a given data distribution $p(y_{\sigma_{0}})$
into a sample of Gaussian noise, denoted $y_{\sigma_T} \sim \mathcal{N}(0, 1)$, where $\sigma_t$ is the noise schedule. The schedule $\sigma_t$ is an ordered list of equal-spaced values $\in [0, 1]$, where $\sigma_0=0$, and $\sigma_T=1$. 
To do so, we propose to rely on $v$-objective diffusion~\cite{ddim}, which defines the intermediate forward diffusion steps as:
\begin{equation}
\fix{y_{\sigma_t} = \alpha_{\sigma_t}y_{\sigma_0} + \beta_{\sigma_t}\epsilon}\label{eq:v_objective}
\end{equation}

where $\epsilon \sim \mathcal{N}(0, 1)$. To calculate $\alpha_{\sigma_t}$ and $\beta_{\sigma_t}$ for a given $t$, 
we define $\phi_t := \frac{\pi}{2}\sigma_t$, and we obtain the corresponding trigonometric values by: $\alpha_{\sigma_t} := cos(\phi_t)$ and $\beta_{\sigma_t} := sin(\phi_t)$.

Intuitively, the diffusion process iteratively adds a small amount of noise to $y_{\sigma_0}$ until reaching an isotropic Gaussian sample 
at $\sigma_T$, while a neural network is trained to perform the reverse operation.
Thereby, the model approximates distribution $\hat{p}(y_{\sigma_0})$, from where observations $\hat{y}_{\sigma_0}$ can be sampled by iteratively denoising random samples of Gaussian noise.

However, this is an intricate problem to learn, which notably benefits from extensive training processes and large datasets. To alleviate such limitations, we propose to operate the diffusion process on a musically and vocally rich latent space instead of directly generating waveforms. \fix{The motivation is to leverage the pre-learnt knowledge from the latent space of the neural codec to learn the task faster and more efficiently.}

\subsection{Latent encoder}\label{sec:intro_codec}
We propose to apply the $v$-diffusion process on the latent space of EnCodec~\cite{defossez2022highfi}. This is a neural codec that provides flexibility, high-efficiency, and pre-trained weights. Alternative codecs do not provide code or model checkpoints, are not optimized for vocal data or present limited performance, or were found not efficient enough for an acceptable sampling time.
\fix{EnCodec leverages a convolutional encoder-decoder architecture, where the encoder maps the input audio into the latent space using convolutional layers augmented with LSTM units, capturing both fine-grained and long-term temporal dependencies. The latent representation is quantized using residual vector quantization (RVQ)~\cite{Vasuki2006ARO}, which discretizes the continuous latent space into a finite set of codebooks. This enables efficient compression with minimal information loss.}

\fix{The decoder reconstructs the audio signal from the quantized latent, mirroring the architecture of the encoder. To improve output fidelity, EnCodec incorporates a multi-scale spectrogram adversarial loss, which minimizes artifacts and enhances perceptual quality by aligning reconstructed audio with the target distribution.
Conveniently, EnCodec is trained on large collections of solo speech, music including vocal recordings, and general audio. The total amount of training time is $\approx15$k hours. For all these reasons, we hypothesize that EnCodec may potentially provide rich embeddings for our singing voice separation, and allow for efficient and perceptually acceptable decoding.}

Let $E$ be the encoder in EnCodec which encodes an audio signal such that $E: x \in \mathbb{R}^{A \times S} \rightarrow X \in \mathbb{R}^{F \times D}$,
where $A$ are audio channels, $S$ length of audio in samples, $F$ are feature channels of the encoded representation and $D = \frac{S}{cf}$ the time dimension of the latent vector, defined by compression factor $cf$ of the latent encoder.
EnCodec may apply further compression using an RVQ-based quantizer $Q$, which quantizes $X$ into a vector of discrete values
$Q: X \in \mathbb{R}^{F \times D} \rightarrow X_q \in \mathbb{Z}^{n_q \times D}$,
where $n_q$ is the codebook size. To convert the compressed representations back to the audio domain, we first denote the de-quantization process as
$dQ:X_q \in \mathbb{Z}^{n_q \times D} \rightarrow \hat{X} \in \mathbb{R}^{F \times D}$, and finally we use decoder
$dE:\hat{X} \in \mathbb{R}^{F \times D} \rightarrow \hat{x} \in \mathbb{R}^{A \times S}$.

We conduct a study on the robustness of EnCodec in reconstructing encoded samples that have been contaminated with Gaussian noise.
%
Such investigation is crucial to identify potential problems due to the use of EnCodec embeddings for latent diffusion. 
We use the vocal tracks in musdb18hq~\cite{musdb}, and compute Signal-to-Distortion ratio (SDR)~\cite{metrics} between the reconstructed signal and the reference.

A complete forward pass of EnCodec involves functions $E$, $Q$, $dQ$ and $dE$, in that order. We contaminate the inner representations at different points, and evaluate how the system behaves when reconstructing the signals. We adapt EnCodec such that $Q$, and subsequently $dQ$, may be skipped, and thus, the latent is never quantized.
Let $std(x)$ be the standard deviation of a signal $x$. Before contaminating the latents, we match the deviation of the noise to that of the latent, therefore the noise is spread through the entire latent.

Let $SDR(x, \hat{x})$ be the SDR metric between a vocal signal $x$ and its reconstruction $\hat{x}$. The performed experiments are:

\input{tables/robustness_fancy}

\begin{enumerate}
    \item \textbf{Identity}: EnCodec is not involved, noise is defined in the time domain and denoted $\epsilon \sim \mathcal{N}(0, std(x)) \in \mathbb{R}^{A \times S}$, and is directly applied on the waveform.
    \begin{equation}
        SDR(x, x+\epsilon)
    \end{equation}
    \item \textbf{No quantization~(NQ)}: quantization is not used, and noise $\varepsilon \sim \mathcal{N}(0,std(X)) \in \mathbb{R}^{F \times D}$ is applied on the non-quantized latent $X$ and is then decoded.
    \begin{equation}
        SDR(x, dE(E(x)+\varepsilon))
    \end{equation}
    \item \textbf{Before quantization~(BQ)}: noise $\varepsilon$ is applied after encoding but before quantization, hence we again contaminate the non-quantized latent.
    \begin{equation}
        SDR(x, dE(dQ(Q(E(x)+\varepsilon))))
    \end{equation}
    \item \textbf{After quantization~(AQ)}: noise $\epsilon_q \sim \mathcal{N}(0, std(X_q))\in \mathbb{Z}^{n_q \times D}$ is applied after encoding and quantizing, therefore we contaminate the quantized latent $X_q$. 
    \begin{equation}
        SDR(x, dE(dQ(Q(E(x))+\varepsilon_q)))
    \end{equation}
\end{enumerate}

See Table~\ref{tab:robustness} for a report on the robustness test.
The results suggest that EnCodec is robust to noise in the non-quantized latent for deviation $\in [1^{-6}, 0.1]$, both when quantizing after adding the noise (\textbf{BQ}) and when skipping quantization (\textbf{NQ}). This indicates that residual noise from the generative sampling process may be further suppressed. For noise with deviation $1$, we report $\approx5$ dB of reconstruction error, albeit we do not expect to decode samples with such level of contamination. The quantization implies loss of $\approx1.1$ dB of SDR. %
We observe that the quantized latent is more sensitive to noise (\textbf{AQ}).
In any case, handling discrete codes necessitates caution, either by accommodating the diffusion process or by mapping these into a continuous representation, adding another layer of complexity. For these reasons, we skip quantization $Q$ and apply the diffusion process on continuous latents $X \in \mathbb{R}^{F \times D}$. These yield substantial compression and competitive reconstruction.

Note that the \textbf{NQ} test for $0$-noise reduces the reconstruction score by an order of magnitude. This suggests that EnCodec-based diffusion is limited by the reconstruction error of such. Being aware that such impact may be minimized by, for instance, fine-tuning the codec using examples of the target distribution~\cite{hwang2023hiddensinger}, we leave such research line for future work.

\subsection{Training algorithm}
Let $X_{\sigma_{0}} \in \mathbb{R}^{F \times D} \sim p(X_{\sigma_0})$ be a latent vector corresponding to a solo singing voice signal, at diffusion time-step $t=0$.
In practice, $v$-objective diffusion models are trained to estimate \textit{velocity} $v_{\sigma_t}$ This term serves as a parameterization that combines aspects of both the original data and the noise, facilitating more stable and efficient training. The target $v_{\sigma_t}$ is formally computed as:
\begin{equation}
    \fix{v_{\sigma_t} = \alpha_{\sigma_t}\varepsilon - \beta_{\sigma_t}X_{\sigma_0}}
\end{equation}
where $\varepsilon \sim \mathcal{N}(0, 1) \in \mathbb{R}^{F \times D}$.
We now refer to Eq.~\ref{eq:v_objective} to compute the forward diffusion input variable $X_{\sigma_t}$:
\begin{equation}
\fix{X_{\sigma_t} = \alpha_{\sigma_t}X_{\sigma_0} + \beta_{\sigma_t}\varepsilon}\label{eq:v_objective_2}
\end{equation}
%

Finally, let $C \in \mathbb{R}^{F \times D}$ be the conditioning signal. 
Unconditioned diffusion models are only able to sample random observations from the approximated distribution $\hat{p}(X_{\sigma_0})$.
However, DDPM trained in an unconditioned manner can still be guided by sampling from the posterior using a conditioning signal to manipulate the diffusion process~\cite{mariani2023multi}.
Interestingly, recent studies have shown that conditioning during both training and inference enhances the system capability to address inverse problems~\cite{diffwave,universe_serra,controlnet}.
In the MSS problem, users expect a well-defined output.
Therefore, it is crucial to guide the diffusion system toward the desired singing voice observation.
For that reason, we provide the model with conditioning signal $C \in \mathbb{R}^{F \times D}$, which corresponds to the latent vector of the mixture signal paired with the singing voice $X_{\sigma_0}$.

We may now estimate the velocity objective $v_{\sigma_t}$ using neural network $m$ with parameters $\theta$:
\begin{equation}
    \fix{\hat{v}_{\sigma_t} = m_{\theta}(X_{\sigma_t}, \sigma_t, C)}
\end{equation}
and formally, the system is trained by minimizing the following expression~\cite{ddim}:
\begin{equation}
\fix{\mathbb{E}_{t\sim[0, T],\sigma_t,X_{\sigma_t}}
\big[||\hat{v}_{\sigma_t} - v_{\sigma_t}||_{2}^{2}\big]}\label{eq:elbo}
\end{equation}
\fix{where $\mathbb{E}$ stands for the expectation. This is known as the evidence lower bound (ELBO) and allows the problem to be tractable using the reparametrization trick~\cite{diffusion-first}}. 

In essence, the $v$-objective diffusion system is trained to approximate the data distribution corresponding to solo vocals encoded in the latent space of EnCodec, while mixture conditioner $C$ is used to guide the sampling process towards the desired observation. We hypothesize that this approach may contribute to generating examples with low interference.

\begin{figure}[t!]
 \centerline{
 \includegraphics[width=0.9\columnwidth]{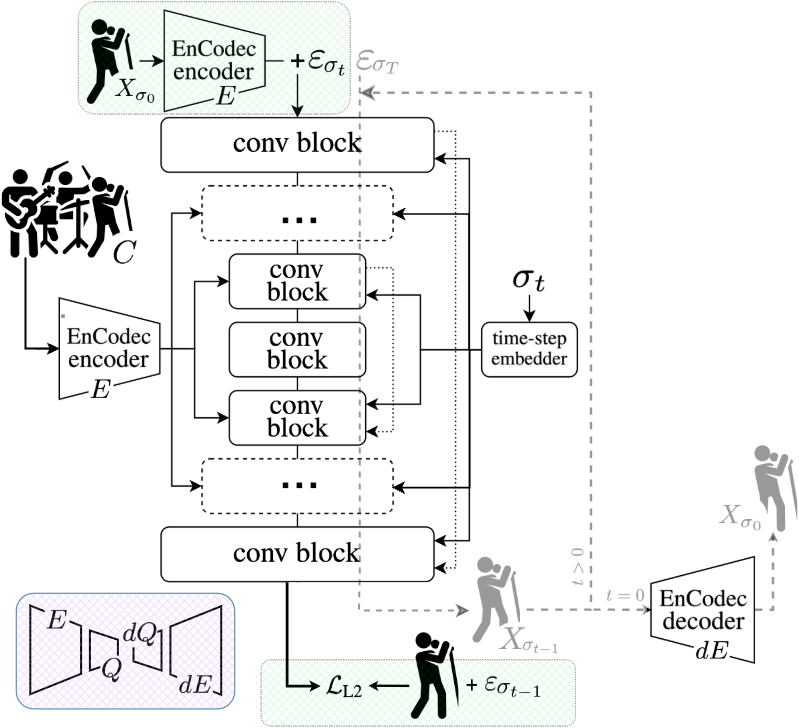}}
 \caption{\textbf{Complete system.} The boxed elements in yellow are only active during training, the rest of the system is used during training and inference. To train, we encode the target vocals and apply the forward diffusion process. The generator network is optimized to estimate a slightly denoised version of the noised target, given step $t$. The corresponding mixture is also encoded and used as conditioner. The gray dashed lines depict the inference stream, which is run for $T$ steps, to convert a Gaussian sample to the vocals contained in the conditioning mixture. Skip connections are depicted in thin, dashed arrows. Boxed in purple, the structure and nomenclature for EnCodec.}
 \vspace{-0.3cm}
 \label{fig:system}
\end{figure}

\subsection{Sampling process}\label{sec:sampling}
The sampling process iteratively denoises a sample of Gaussian noise to get to an observation of the approximated target distribution $\hat{p}(X_{\sigma_0})$. This is depicted using a gray dashed line in Figure~\ref{fig:system}. Previous audio generation works have used the Denoising Diffusion Implicit Models (DDIM) sampler~\cite{salimans2022progressive}, which provides an acceptable balance between sampling steps and generation quality~\cite{schneider2023mousai}.

Let us define the complete set of operations in a single DDIM sampling step. We first run inference for a given $t$, estimating velocity $\hat{v}_{\sigma_t} \in \mathbb{R}^{F \times D}$:
\begin{equation}
\fix{\hat{v}_{\sigma_t} = m_{\theta}(X_{\sigma_t}, \sigma_t, C)}
\end{equation}
Now, having predicted velocity $\hat{v}_{\sigma_t}$ and the forward diffusion variable $X_{\sigma_t}$ at hand, we may proceed to compute:
\begin{gather}
\fix{\hat{X}_{\sigma_0} = \alpha_{\sigma_t}X_{\sigma_t} - \beta_{\sigma_t}\hat{v}_{\sigma_t}} \\
\fix{\hat{\varepsilon}_{\sigma_t} = \beta_{\sigma_t}X_{\sigma_t} + \alpha_{\sigma_t}\hat{v}_{\sigma_t}}
\end{gather}
which intuitively, $\hat{X}_{\sigma_0} \in \mathbb{R}^{F \times D}$ corresponds to the predicted latent of the target sample at $t=0$, and $\hat{\varepsilon}_{\sigma_t} \in \mathbb{R}^{F \times D}$ to the also predicted Gaussian noise at current $t$. For $t\approx T$, the predicted $\hat{X}_{\sigma_0}$ is not yet a satisfactory generation. However, as $t$ gets closer to $0$, the predicted samples are refined and further adhere to the conditioning signal.

Finally, the input for the next sampling step is computed as:
\begin{equation}
    \fix{\hat{X}_{\sigma_{t-1}} = \alpha_{\sigma_{t-1}}\hat{X}_{\sigma_0} + \beta_{\sigma_{t-1}}\hat{\varepsilon}_{\sigma_t}}
\end{equation}

$T$ determines the granularity of the sampling process, however, large $T$ does not necessarily imply better sampling. In fact, strongly conditioned systems may require or even perform better using small $T$ values~\cite{universe_serra}.

\subsection{Model architecture}\label{sec:network}
\subsubsection{Generator U-Net}
\begin{figure}[t!]
 \centerline{
 \includegraphics[width=0.575\columnwidth]{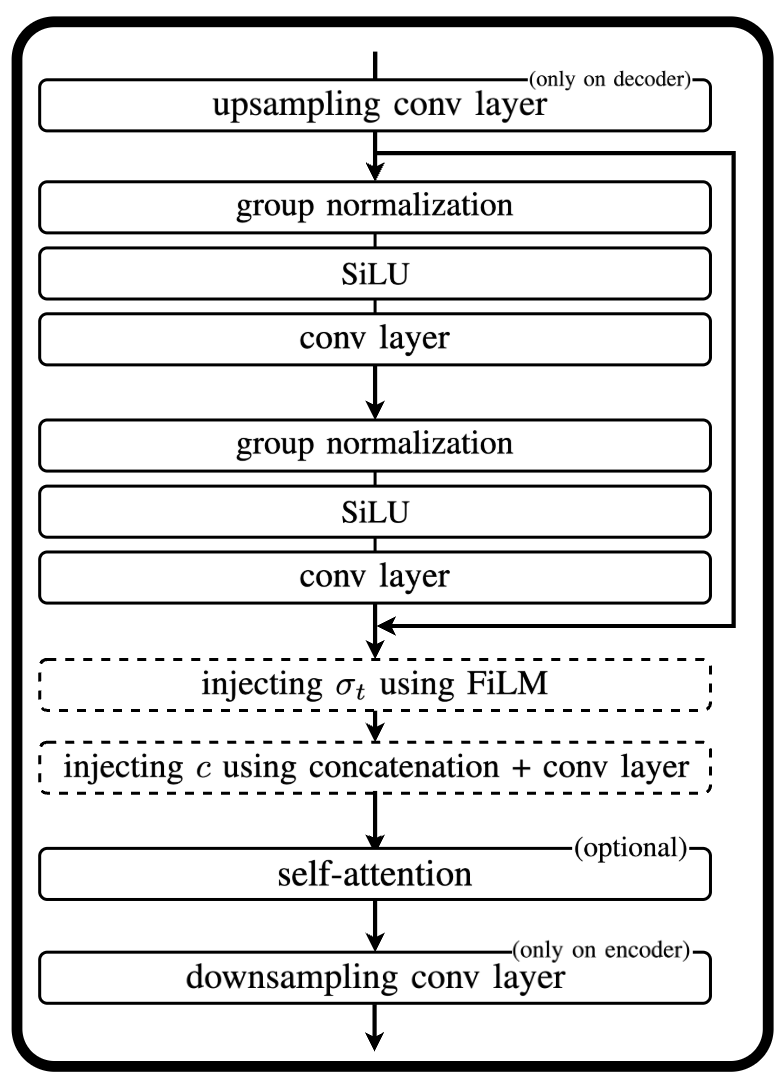}}
 \caption{U-Net block of the generator network.}
 \vspace{-0.3cm}
 \label{fig:block}
\end{figure}

U-Net architectures have demonstrated success for score-based diffusion~\cite{diff_beat_gans}. \fix{We use a convolutional U-Net composed of residual blocks, which are illustrated in Figure~\ref{fig:block}. These comprise two sequential 1D convolutional layers, each preceded by group normalization and SiLU activation function.
We propose to employ 1D convolutions to effectively capture temporal dependencies in the EnCodec embeddings, processing each feature vector independently without imposing artificial spatial correlations. Using 2D convolutions to process the learned embeddings may cause the network to prioritize signal reconstruction over feature processing. In fact, preliminary experiments using 2D convolutions resulted in more interferences. Directly compressing the EnCodec embeddings using a 1D U-Net results in cleaner vocals while using less parameters and therefore, enhancing computational efficiency.}

\fix{Residual connections help preserving structural and contextual information from the input, facilitating more effective gradient flow, and improving stability.
We optionally insert time-wise self-attention to enhance the ability to capture long-range dependencies. 
This enables the U-Net to balance local feature extraction with the global context, which is important for MSS.
Finally, a $1 \times 1$ strided convolution is used to downsample in the encoder, and a transposed convolution is used to upsample in the decoder.}

The diffusion time-step $\sigma_t$ is projected into a 1024-channeled random Fourier feature embeddings, which are processed through a 3-layer multi-layer perceptron (MLP) with GELU activations. The resulting embedding is incorporated into the model via FiLM layers. Specifically, the embedded diffusion step is injected after the two convolutional layers and the residual sum, as depicted in Figure~\ref{fig:block}.

\subsubsection{Conditioning processing and injection}\label{sec:conditioner}
We first encode the conditioning signal: $C = E(c)$. We \fix{use bilinear interpolation} to match the inner layers of the generator at each level of depth.
We insert $C$ by concatenating with the network layers through the feature channels, followed by a $1 \times 1$ convolutional layer to \fix{dynamically merge the features}.
$C$ is not necessarily injected to all generator levels, as seen in Figure~\ref{fig:system}. More injections imply stronger conditioning, although more trainable parameters are needed.

\subsubsection{A two-stage training approach}
Despite having the pre-trained EnCodec, one may argue that the encoder that processes $c$ is not optimized to separate. Thus, the separation effort is carried out entirely by the generator, which is already in charge of the generation. 
To maintain stability while fine-tuning the conditioner for our problem, we split the training process in two stages: \textbf{(1)}~we load the pre-trained conditioning encoder, freeze it, and train the system for a number of steps, \textbf{(2)}~we unfreeze the encoder and continue training the entire system at a small learning rate.
The encoder and decoder used to encode $X_{\sigma_0}$ prior to the forward diffusion process, depicted at the top and bottom of Figure~\ref{fig:system}, are kept frozen at all times.

\section{Experiments}\label{sec:experiments}

\subsection{Experimental setup}\label{sec:setup}
The generator U-Net is composed of $7$ levels of depth, with feature channels that start at $128$ -- note that EnCodec latents are sized $128$ -- and are duplicated every two levels until reaching size $1024$ in the bottleneck. 
We inject the conditioning mixture to all levels but the three first, where less feature channels are available, aiming at reducing the flow of accompaniment directly to the output.
During training, we use a time context of $\approx13$s, corresponding to $602000$ samples,
which is compressed down to $2048$ by the neural codec.
The generator compresses the time dimension by a factor of $2$ to all levels but the foremost two, reaching length of $32$ in the bottleneck. 
The system has 115.9M parameters.

We use EnCodec pre-trained at 48kHz, since a 44.1kHz model is not available.
We resample before running forward diffusion, and after sampling during inference.
%
We normalize the latents ensuring that $\text{std}(X_{\sigma_t})<=1$, otherwise it may negatively effect proper optimization of the diffusion U-Net, leading to inconsistent amplitude of the decoded waveform and noisy sampling.
Moreover, by default, EnCodec processes audio signals in overlapped segments, which are then combined through overlap-add. Preliminary results show that decoding a single stream reduces hallucinations and generation artifacts. Therefore, discard chunk-wise processing.
%

The first training stage runs for 600k steps and takes about one week in four A100 GPU. We use effective batch size of $32$, and AdamW optimizer with a weight decay of $1*10^{-3}$ and initial of learning rate $2*10^{-4}$. 
Next, we unfreeze the conditioner encoder, and continue the training for 400k steps, using a learning rate of $5*10^{-5}$. 
%
%
We integrate several data augmentation techniques during training: polarity inversion and channel flipping, pitch shifting an stem remixing.
To sample, we use $T=50$, forwarding the mixture in chunks of $602000$ samples with a 20\% overlap.
%
%
%
We use musdb18hq~\cite{musdb} for training, aiming at studying the data-efficiency of the proposed system.
Note that related diffusion works rely on larger collections e.g. $1500$~\cite{universe_serra}, $2500$~\cite{schneider2023mousai}, or $19500$ h~\cite{evans2024fast}. To evaluate, we use the testing set of musdb18hq.

\input{tables/objective_only}

\subsection{Objective evaluation}
\camred{Recent studies have pointed out the limitations in using SDR as a sole metric for assessing the perceptual quality of separation systems~\cite{guso2022loss, cano2016evaluation}.
Moreover, it has been shown that SDR may not be able to fully represent the performance of generative approaches~\cite{pritish_sep}. The stochasticity in the sampling process of these systems may produce phase mismatch and signal misalignment with the ground-truth, which although unperceivable at naked ear, penalize the scores.
Moreover, the presence of a latent encoder may exacerbate phase discrepancy, further complicating objective assessment. In fact, existing LDM for separation exclude SDR to evaluate~\cite{msdm_ldm}.}

\camred{Therefore, we assess our generations on objective quality measures for speech synthesis and music generation.
We report pitch error computed as the mean-squared error of the logarithmic pitch values computed using a monophonic pitch tracker~\cite{hwang2023hiddensinger}. We also report Log-spectral distance (LSD) and Mel-spectrogram Mean Absolute Error (Mel-MAE), both aiming at computing signal quality independently from the phase as suggested in~\cite{liu2023audioldm, hwang2023hiddensinger, msdm_ldm}.
Finally, we compute Perceptual Evaluation of Speech Quality (PESQ)~\cite{pesq} to measure intelligibility, being aware that phase discrepancies and presence of leaked accompaniment may have a direct impact on the scores.
We compute the metrics on the voiced regions~\cite{hwang2023hiddensinger}.}

\camred{We compare our model, LDM-dmx, against source-only InstGlow~\cite{instglow} and MSDM~\cite{mariani2023multi}. Since no trained vocal model for MSDM is publicly available, we train it using musdb18hq following the instructions in the official repository.
MSG-LD~\cite{msdm_ldm} is excluded, as it employs a latent encoder tailored to instrumental music; adapting it for vocals would require substantial structural changes, making comparison infeasible.}

\camred{For reference purpose only,} we include two non-generative models in the experiments: Band-Split RNN (BS-RNN)~\cite{Luo2022}, which operates on spectrograms, and Hybrid Demucs (H-Demucs)~\cite{defossez2021hybrid}, which combines information from the time and time-frequency domains to leverage the strengths of each. \camred{These systems are trained with access to all available sources, which is advantageous for the model optimization~\cite{instglow}}.

\subsection{Listening test}
Given the recurrent inconsistency between objective separation metrics and perceptual quality, the interest for perceptual assessment has raised in the MSS community~\cite{sdx_2023}. In a generative context, perceptual experiments are especially useful to complement objective measures.
%
We run a perceptual test following the ITU-T P.835 recommendation. We rely on the MUSHRA framework~\cite{schoeffler2018webmushra}, and gather Mean Opinion Scores (MOS) between 1 and 5, being 5 the maximum score, on the following aspects:
\textbf{(1)}~intelligibility, \textbf{(2)}~interferences, and \textbf{(3)}~artifacts~\cite{pritish_sep}. \fix{In the context of singing voice generation, it is crucial to ensure that the output does not sing nonsensical lyrics. To prioritize this, intelligibility is assessed first in our test, preventing participants from becoming overly familiar with the lyrics while evaluating other aspects.}

The participants are given the mixture as reference.
\camred{For each example, four randomly ordered and unnamed stimuli are shown. These include LDM-dmx, BS-RNN (time-frequency based), H-Demucs (time plus time-frequency based), and MSDM (the best performing generative baseline).}
\camred{We do not include the IRM as it is perceptually close to the reference vocal stimuli which is not shown in the test.}

We separate the entire musdb18hq test set, randomly sample 6 recordings, and then randomly select a 30s chunk for each, checking it is voiced on more than the 50\% of its duration.

\input{tables/perceptual_only}

\section{Results}\label{sec:results}
\camred{Table~\ref{tab:objective_results} reports objective evaluation of our system against generative and non-generative baselines.
The LSD and Mel-MAE scores suggest that our system achieve competitive performance in modeling the magnitude spectral components, independently from the phase discrepancies.
In terms of log pitch error, LDM-dmx outperforms the generative baselines and levels the non-generative systems, suggesting competitive generation of the vocal harmonic components and low interference from the accompaniment.}

\camred{While leading against the generative baselines, the margin between LDM-dmx and the deterministic systems on PESQ is more significant than for other metrics. The discrepancy in the high-frequency range in combination with potential generation artifacts may penalize this speech metric which focuses on synthesis quality and intelligibility.
Note that generative modeling followed by latent decoding is strongly prone to phase discrepancy when addressing inverse problems. Albeit often unperceivable at naked ear, this may directly have a negative impact on the PESQ scores.}

See the perceptual report in Table~\ref{tab:perceptual_results}.
A total of 15 participants took the test. 
To illustrate the level of participant agreement, we report the Confidence Intervals (CI) with $\alpha=0.05$.

\camred{Our system is leading against the generative baseline. However, the non-generative systems outperform the generative in terms of artifacts, albeit LDM-dmx reduces the signal quality gap to a difference of $\approx0.6$ points.
In this same line, LDM-dmx is scoring $\approx0.7$ points lower than both BS-RNN and H-Demucs on the intelligibility scale, matching what the PESQ scores suggest in Table~\ref{tab:objective_results}. However, the perceptual scores suggest that although non-generative models still output clearer vocals, our system mostly generates understandable lyrics.
However, the combination of generation and decoding artifacts
may be shadowing particular words or sounds.}

In terms of interference removal, our system levels BS-RNN and outperforms H-Demucs and MSDM. \fix{This may be} due to the generator network being trained to approximate the distribution of clean singing vocals, which in combination with the proposed training scheme, enables the generation of a notably clean output. The pre-learnt knowledge of solo vocals in EnCodec is also arguably beneficial. \camred{The isolation capability is likely having a positive impact on the scores in Table~\ref{tab:objective_results}.}

\camred{The latent diffusion framework and model size enable the fastest inference among generative systems
and performs comparably to H-Demucs.} \camred{Latent diffusion allows also forwarding wider context, an important factor for MSS.}
Despite these gains in efficiency, \camred{our objective and perceptual evaluation suggest that the generations may potentially include artifacts that are perceptible at naked ear.} 
Fine-tuning the latent encoder with extensive collections of vocal recordings may potentially enhance quality and mitigate artifacts~\cite{hwang2023hiddensinger}. 
%
%
Furthermore, we hypothesize that LDM-dmx would exploit access to supplementary training data, which could even be synthetically generated, as the system solely requires pairs of corresponding isolated vocals and mixtures.

\section{Conclusions}
Diffusion models are emerging as a promising alternative to frequency masking or transformation approaches for music source separation, addressing limitations such as the source overlap in music or predicting the phase of complex spectrograms. 
However, diffusion models require large multi-stem datasets and are computationally intensive for training and sampling.
To make such tools more accessible to music practitioners, we investigate the use of latent diffusion for source-only separation of the singing voice. Our approach significantly reduces training requirements, and it is trained using only open multi-stem data.
%
The system is optimized to generate separated singing vocals conditioned on music mixtures.
We evaluate our model on objective speech synthesis and audio generation quality metrics. Furthermore, we run a listening test to collect human assessment and contrast the objective measures, offering a practical perspective on the performance of on our system, which we frame best on a creative or educational musical context.
The results suggest that our system outperforms the generative baselines and levels the non-generative models on a list of spectral audio quality measures. We achieve competitive interference removal, although perceptible high-frequency artifacts, added to the reconstruction error of the neural decoder, are present.
\camred{We envision further work toward refining the source quality, reduce artifacts, and extend the approach to other sources.}

\bibliographystyle{IEEEtran}
\bibliography{refs}

\end{document}

%% file: tables/robustness_fancy.tex
\begin{table}[t!]
\caption{\textbf{Noise robustness of EnCodec.} Bottom row provides number of feature channels of the corrupted vector. Results in dB.}
\vspace{-0.1cm}
\centering
\renewcommand{\arraystretch}{1.1}
\begin{tabular}{lcccc}
\toprule
\textbf{Std. dev. of $\epsilon$} & \textbf{Identity} & \textbf{NQ} & \textbf{BQ} & \textbf{AQ} \\
\midrule
\textbf{0.0} & 119.24 & 10.81 & 9.77 & 9.77 \\
\textbf{1e-6} & 96.64 & 10.81 & 9.77 & -0.22 \\
\textbf{1e-3} & 36.67 & 10.81 & 9.77 & -0.43 \\
\textbf{0.01} & 16.67 & 10.81 & 9.77 & -2.47 \\
\textbf{0.1} & -3.32 & 10.74 & 9.72 & -2.80 \\
\textbf{1.0} & -23.32 & 5.51 & 5.57 & -2.85 \\
\midrule
\textbf{Feat. size} $\rightarrow$ & 2 & 128 & 128 & 16 \\
\bottomrule
\end{tabular}
\label{tab:robustness}
\vspace{-0.32cm}
\end{table}

%% file: tables/objective_only.tex
\begin{table*}[t!]
\caption{\textbf{Objective evaluation of our system against the generative baseline, reference non-generative models, and the Ideal-Ratio Mask.} Arrows indicate if higher ($\uparrow$) or lower ($\downarrow$) is better. Following \cite{mariani2023multi}, inference time is computed on a $\approx$12s excerpt using a 12GB A100 GPU. While most metrics are distance measures, the maximum score for PESQ is 4.5.}
\vspace{-0.14cm}
\centering
\renewcommand{\arraystretch}{1.1}
\begin{tabular}{lcccccccc}
\toprule
\textbf{Model} & \textbf{Gen?} & \textbf{\#Param.} & $\mathbf{T}$ (for best results) & \textbf{Inference time (s) $\downarrow$} & \textbf{LSD $\downarrow$} & \textbf{Mel-MAE $\downarrow$} & \textbf{Log-F0 RMSE $\downarrow$} & \textbf{PESQ $\uparrow$} \\
\midrule
\midrule
\textbf{IRM}~\cite{sisec2018} & \xmark & -- & -- & -- & 0.64 & 1.84 & 0.119  & 2.98 \\
\midrule
\textbf{InstGlow}~\cite{instglow} & \cmark & 15M & $150$ & 16.50 $\pm$ 0.12 & 2.17 & 8.40 & 0.268 & 1.17 \\
\textbf{MSDM$_{\text{dirac}}$}~\cite{mariani2023multi} & \cmark & 405M & $150$ & 11.95 $\pm$ 0.14 & 2.08 & 8.21 & 0.369 & 1.16 \\
\textbf{LDM-dmx} & \cmark & 116M & $50$ & 1.74 $\pm$ 0.02 & 1.30 & 6.43 & 0.198 & 1.29 \\

\midrule
\textbf{BS-RNN}~\cite{Luo2022} & \xmark & 36M & -- & (no official code out) & 1.41 & 6.94 & 0.187 & 1.98 \\
\textbf{H-Demucs}~\cite{defossez2021hybrid} & \xmark & 346M & -- & 0.40 $\pm$ 0.09 & 1.01 & 5.11 & 0.172 & 1.82 \\ 
\bottomrule
\end{tabular}
\label{tab:objective_results}
\vspace{-0.32cm}
\end{table*}

%% file: tables/perceptual_only.tex
\begin{table}[t!]
\caption{\textbf{Perceptual MOS evaluation of perceptual separation quality.} The columns are placed in the same order as the aspect to evaluate is shown to the participants.}
\vspace{-0.09cm}
\centering
\renewcommand{\arraystretch}{1.2}
\begin{tabular}{lccc}
\toprule
\textbf{Model} & \textbf{Intelligibility $\uparrow$} & \textbf{Isolation $\uparrow$} & \textbf{Sig. quality $\uparrow$} \\
\midrule
\midrule
\textbf{LDM-dmx} & $2.76_{[2.59,2.93]}$ & $4.04_{[3.86,4.22]}$ & $3.43_{[3.24,3.62]}$ \\
\textbf{MSDM$_{\text{dirac}}$}~\cite{mariani2023multi} & $2.40_{[2.14,2.66]}$ & $1.98_{[1.73,2.22]}$ & $2.80_{[2.55,3.06]}$ \\
\midrule
\textbf{BS-RNN}~\cite{Luo2022} & $3.40_{[3.19,3.62]}$ & $4.09_{[3.90,4.28]}$ & $4.00_{[3.80,4.18]}$ \\
\textbf{H-Demucs}~\cite{defossez2021hybrid} & $3.45_{[3.25,3.66]}$ & $3.62_{[3.41,3.83]}$ & $4.03_{[3.83,4.22]}$ \\
\bottomrule
\end{tabular}
\label{tab:perceptual_results}
\vspace{-0.32cm}
\end{table}